\documentclass[aps,prb,print,groupedaddress,preprint,showpacs]{revtex4-1}
\usepackage{amssymb}
\usepackage{textcomp}					% sonderzeichen
\usepackage{graphicx}
\usepackage{epstopdf}
\usepackage{dcolumn}
\usepackage{bm}
\usepackage{array}
\usepackage{tabto}
\usepackage{booktabs}
\usepackage[T1]{fontenc}
\usepackage{color}
\newcommand{\ra}[1]{\renewcommand{\arraystretch}{#1}}

\begin{document}
\graphicspath{{LumiFigures/}}
%\title{The gallium problem for high mobility GaAs/AlGaAs heterostructures}

\title{Photoluminescence and the gallium problem for highest-mobility GaAs/AlGaAs-based 2d electron gases}
%\author{Several Authors}

%\author{W. Wegscheider}
%\affiliation{ Solid State Physics Laboratory, ETH Zurich, 8093 Z\"urich, Switzerland}

\author{F. Schl\"apfer, W. Dietsche, C. Reichl, S. Faelt and W. Wegscheider}
%\address{ Solid State Physics Laboratory, ETH Z\"rich, 8093 Z\"urich, Switzerland}
\affiliation{ Solid State Physics Laboratory, ETH Z\"urich, 8093 Z\"urich, Switzerland}

%\affiliation{ Max Planck Institute for Solid State Research, 70569 Stuttgart, Germany}

\date{\today}

\begin{abstract}
The quest for extremely high mobilities of 2d electron gases in MBE-grown heterostructures is hampered by the available purity of the starting materials, particularly of the gallium. Here we compare the role of different Ga lots having nominally the highest possible quality on the mobility and the photoluminescence (PL) of modulation doped single interface structures and find significant differences. A weak exciton PL reveals that the purity of the Ga is insufficient. No high mobility can be reached with such a lot with a reasonable effort. On the other hand, a strong exciton PL indicates a high initial Ga purity, allowing to reach mobilities of 15 million  (single interface) or 28 million $cm^2/Vsec$ (doped quantum wells) in our MBE systems. We discuss possible origins of the inconsistent Ga quality. Furthermore, we compare samples grown in different MBE systems over a period of several years and find that mobility and PL is correlated if similar structures and growth procedures are used.\\
Keywords:
A3. Molecular beam epitaxy,
B2. Semiconducting III-V materials,
B3. Heterojunction semiconductor devices

\end{abstract}

%\pacs{73.43.-f,73.43.Lp,73.43.Fj,77.65.Dq} 

\maketitle
{\tiny }

\section{Introduction}

  The development of molecular-beam epitaxy (MBE) has enabled experimenters to produce semiconductormaterials consisting of atomically precise layers of different chemical composition also called heterostructures.\cite{HeniniBook} In this process, the starting materials are heated until  molecular or atomic beams are evaporated which react on the surface of a single-crystalline substrate to form the desired structure. Devices made from these heterostructures have found applications in very different fields of physics.\cite{2010Schlom} Here, we concentrate on heterostructures based upon specific III-V semiconductors, i.e. GaAs and Al$_{x}$Ga$_{1-x}$As where two-dimensional electron (or hole) gases can be found at an interface or in a GaAs quantum well embedded in Al$_{x}$Ga$_{1-x}$As layers.\cite{1988Ploog} The electron mobilities in such systems may reach several ten millions $cm^2/Vs$.\cite{2003Pfeiffer,2009Umansky,2014Manfra,2015DissReichl} These structures turn out to be the testbeds of electron-interaction phenomena in high magnetic fields like the p-wave superconductivity at the filling factor 5/2,\cite{Willett1987} the striped phases at large filling factors,\cite{Fogler1996} and the BCS state in bilayers at the total filling factor one.\cite{Spi2000,Xuting2012}
  
  An enormous experimental effort is required to obtain 2d electron gases (2DEG) with the desired extremely high mobilities,\cite{2003Pfeiffer,2015DissReichl} limiting their availability. One prerequisite is a specialized ultra-high-vacuum MBE-system with elaborate load-locks and an impurity-free substrate-heating system. This can be achieved through extreme care in the design and operation. Another prerequisite is a sufficiently high purity of the starting materials. One needs to consider that the desired impurity concentration in these structures is of the order of $10^{-9}$ (10$^{13}$ atoms $cm^{-3} $) which is two to three orders of magnitude less than the specified impurity levels of the best available materials. Fortunately, the actual purity of the molecular beams will usually be higher than that of the starting materials because Raoult's law states that the vapor pressure of the impurities is reduced by their respective concentration in the starting material.

Unfortunately, the high mobilities are frequently not reproducible. It became clear over the last decade by considering the results in both the authors' and other laboratories that the purity of the gallium is frequently questionable. Using gallium lots with nominally high purity led to low or mediocre 2DEG mobilities. A detailed description of this problem has been presented in a very recent manuscript by Gardner et al.\cite{Gardner2016}

\begin{figure}
 \includegraphics[width=0.6\textwidth]{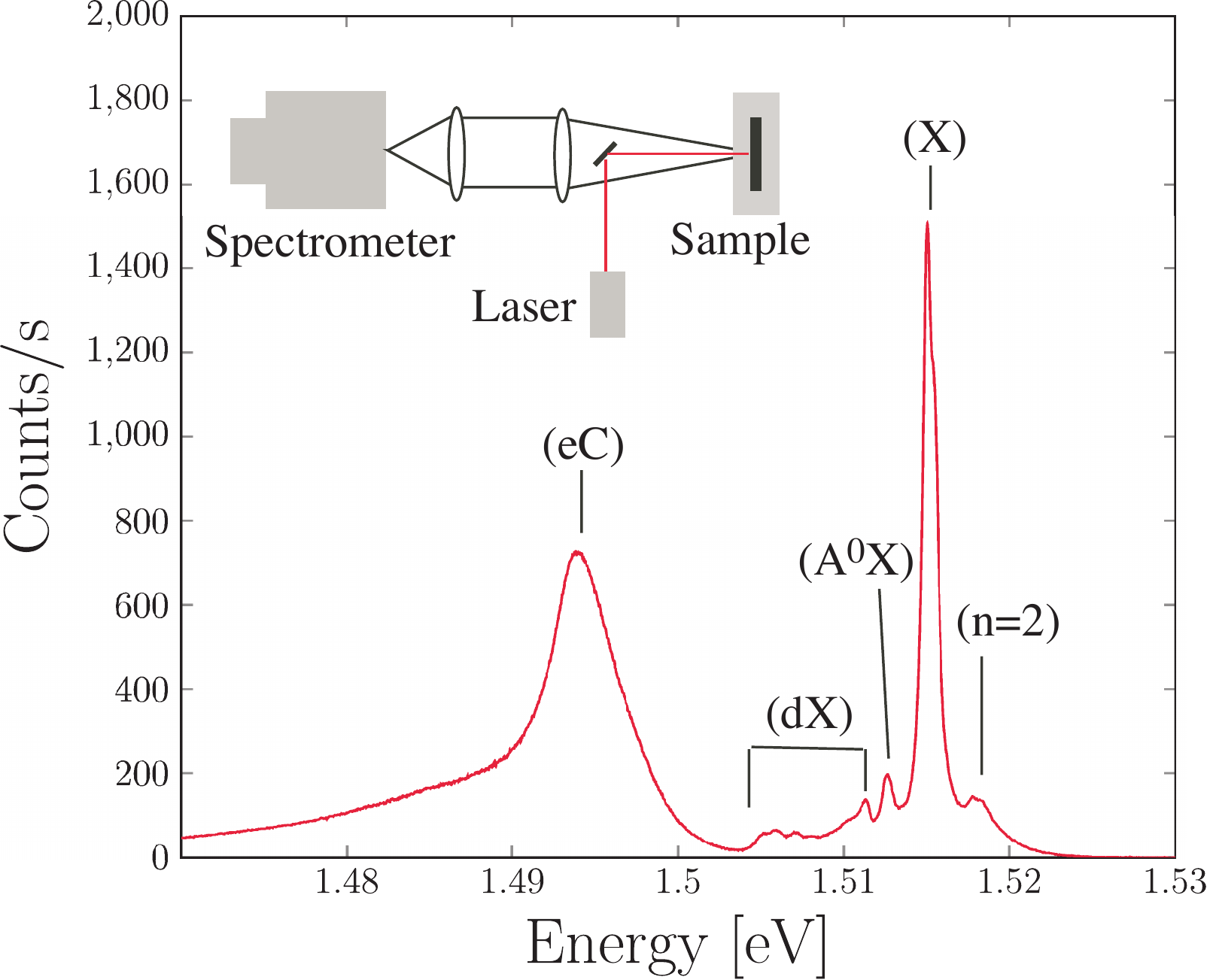}\\ \textbf{\caption{\label{BasicSpectrum}
  \textbf{PL spectrum of a GaAs/AlGaAs heterostrucure containing some impurities. All lines originate from the GaAs. The peaks above 1.5 eV are due to exciton recombinations which originate from (X): free excitons, (n=2): excited excitons, (A$^0$X): excitons bound to neutral acceptors, (dX): excitons bound to neutral defects. At lower energies, the feature (eC) designates a conduction band to carbon acceptor transition. The inset shows the setup of a PL experiment which consists of a HeNe laser exciting the sample and a spectrometer equipped with a CCD camera to analyze the emitted light.}}}
\end{figure}

Presently, most gallium on the market is purified by repeating a partial crystallization of the Ga by which the impurities are reduced. The repetitions are stopped as soon the impurity levels are close to the detection limits of analytical techniques like  GDMS ("Glow Discharge Mass Spectrometry") corresponding to the typically specified 7N quality.
 It turns out that GDMS is of little use for distinguishing "good" and "bad" gallium because no significant differences between the analytical results of the lots leading to poor or high mobilities are found. 
 
 Another analytical tool to characterize impurity levels is the determination of the resistance ratio of the Ga at 300 K and 4 K (residual resistance ratio) although its value is also affected by the crystal structure. Several Ga ingots had been measured inductively by one of the authors.\cite{Werner2012} All showed resistance ratios exceeding 50000 indicating extremely high purity even if the resulting mobilities remained poor.
 
As a result, the quality of a given Ga lot can only be assessed after growing many heterostructures over a period of several months until all trivial sources of contaminations have disappeared from the MBE apparatus and a characteristic mobility can be assessed. Thus, an early analytical technique would be very desirable to save valuable growth time. In this paper, we report on our study of photo-luminescence  of the as-grown wafers which shows great promise as an early indicator of the Ga quality. In particular, we find that Ga lots which nominally have a high purity but lead to low mobilities can be identified at an early stage. It appears that these low-quality Ga lots cause heterostructures with deep levels in the GaAs band-gap which are detrimental to both mobility and luminescence. Before describing our results we note that 
it is not our intention to perform research on the photoluminescence properties of GaAs but to employ this technique for improving the quality of MBE structures, and in particular to give a rapid assessment of the Ga quality.

  \begin{figure}
   \includegraphics[width=0.6\textwidth]{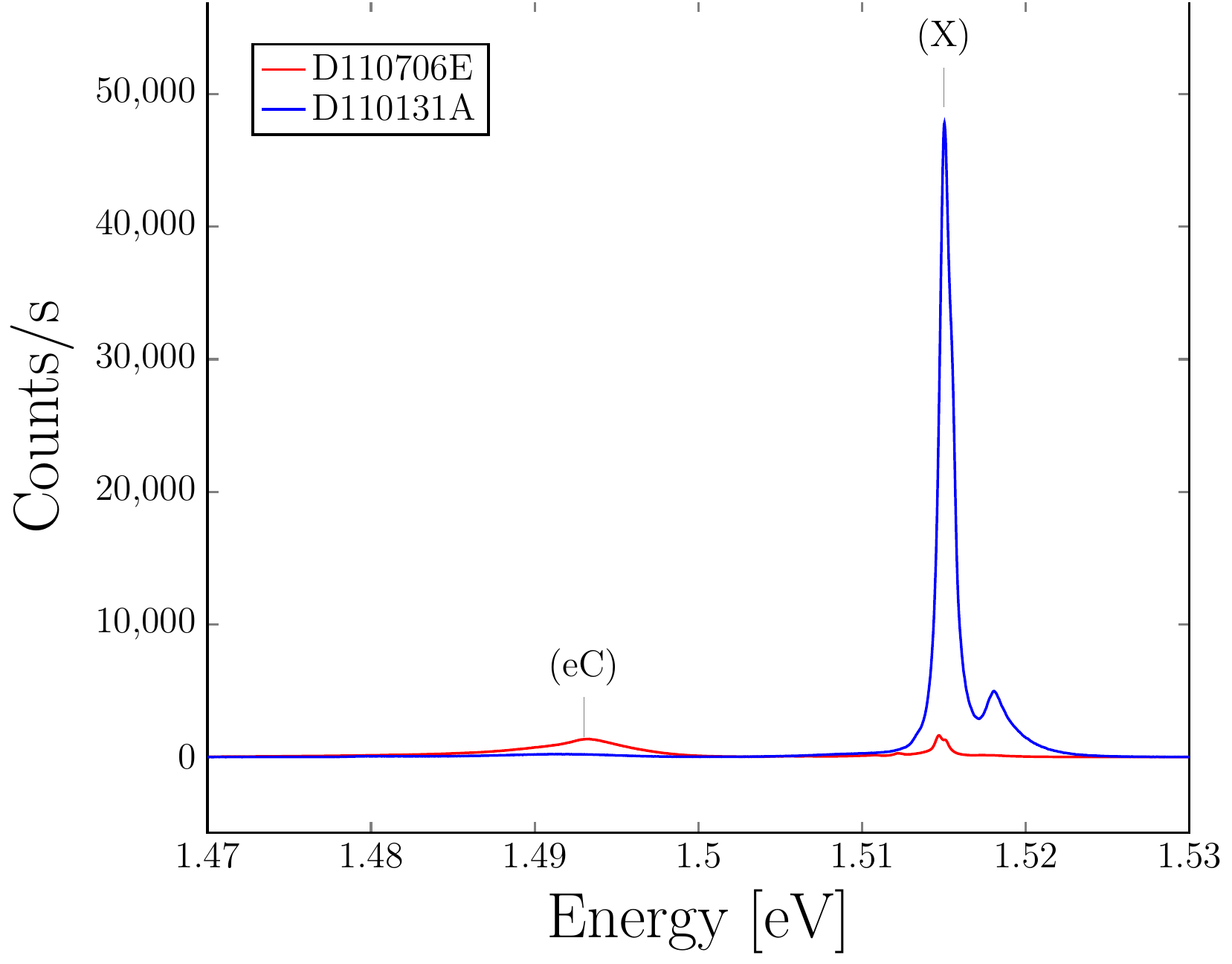}\\
   \textbf{\caption{\label{Fig_4_1}
    \textbf{{GaAs spectra of two MDSI samples grown in the same chamber with two different gallium cells (blue: \textit{Ga4} {\slshape D110131A}, red: \textit{Ga8} {\slshape D110706E}). The only difference between the growth procedures of the two wafers is the choice of the gallium.}}}}
  \end{figure}
  
\section{Characterizing heterostructures by photoluminescence}

Photoluminescence (PL) is a simple and non-destructive measurement technique that allows to get information about impurities in semiconductors. A typical measurement setup is shown as inset in  Fig.~\ref{BasicSpectrum} and 
consists of an excitation laser, the sample in a cryostat, some focussing optics and a spectrometer.

 We use a HeNe laser operating at 632.8 nm (1.9593 eV) to excite the charge carriers in our samples. 
The  power density of the laser is kept constant because it affects the relative intensities in the PL spectra which is used for analyzing and comparing the spectra. We use a laser power of \mbox{83 ${\mu}W$} corresponding to an excitation energy density on the sample of approximately \mbox{0.03 ${W}/{cm^2}$}. The sample is at a temperature of  \mbox{4.7 Kelvin} in an optical continuous flow cryostat. 

The PL light is focused  on the spectrometer (\mbox{RAMONOR} U 1000) equipped with a 600 grooves/mm grating  and a CCD detector (\mbox{Symphony II}, Horiba). The slit size is 150 $\mu$m. Like the excitation power, the detection optics for the luminescent light is unchanged over a measurement series. The reproducibility of the excitation and detection setup is checked frequently by remeasuring the PL of a calibration sample.  We estimate that the laser-power densities do not vary by more than 10 \% between different experiments while the reproducibility of the PL intensity is of the order of 30 \%. 
%D130409A is our calibration sample

Most samples investigated are modulation-doped single interface structures (MDSI) grown on GaAs substrates.  First, a smoothing superlattice is grown, followed by a 1 ${\mu}m$ thick GaAs buffer and a  300 nm thick Al$_{x}$Ga$_{1-x}$As layer.  The x-value is about 0.25.
At the interface between GaAs and  Al$_{x}$Ga$_{1-x}$As a 2DEG forms. The electron concentration of this 2DEG is controlled by the amount of Si $\delta$-doping in the AlGaAs and the distance between the doping and the 2DEG, called spacer, which is between 30 and 70 nm.
 The 2DEG mobility of the samples used in this work varies from 0.1 to 14 million ${cm^2}/{Vs}$. All mobility values are measured at 1.2 K after illumination with a red LED.

In the actual PL experiment, the laser light will be mainly absorbed in the GaAs and to a lesser extend in the AlGaAs. Electrons are excited from the valence band into the conduction band. 
By fast thermalisation and scattering processes, the excited carriers relax quickly towards the band edges where, at low temperatures, they bind to excitons before recombination occurs. In radiative processes, this is accompanied by the emission of photons with the energy $E_{X} = E_g - E_{binding}$ where \textit{E$_g$} is the band gap and  \textit{E$_{binding}$} the exciton binding energy. 

%\begin{equation}

%\label{ground_state_exciton}
%\end{equation}
% The gap and the binding energy of an exciton in Al$_x$Ga$_{1-x}$As depend on the x-value\cite{PL_review}. Tables \ref{bandgap_AlGaAs}  lists the corresponding energies for the directly and the indirect  gap as well as the exciton energies in Al$_x$Ga$_{1-x}$As. \cite{GaAlAs_direct_gap_value,GaAlAs_indirect_gap_value}

The presence of impurities and dopants provides additional recombination channels and leads to separate luminescence lines or to non-radiative recombination without detectable PL.  Furthermore, excitons may be bound to donors or acceptors shifting the exciton energy slightly. These phenomena have already been studied in bulk GaAs in the past and details can be found in Ref.\cite{PL_review} In this work we are interested in the investigation of heterostructures containing a 2DEG. It turns out that these structure have a PL spectrum which largely coincides with that of bulk GaAs. 

 A PL spectrum originating from an AlGaAs/GaAs heterostructure of moderate purity is shown Fig.~\ref{BasicSpectrum}. The peak at 1.5151 ev is assigned to free excitons (X)in good agreement with the 1.5153 eV given in Ref.\cite{PL_review} for excitons in bulk GaAs. In addition, luminescence from the first excited higher state of the excitons is also visible and marked with "n=2".
 
Additonal structures are caused by unintentional impurities. The line at \mbox{1.5126 eV} is assigned to excitons bound to neutral acceptors (A$^0$X) and agrees exactly with the literature values. Excitons bound to neutral donors (D$^0$X) would show up at 1.5143 eV but are not seen in this sample. 
 The next set of lines is spread between 1.504 to 1.512 eV, marked (dX) is ascribed to excitons bound to neutral defects.\cite{PL_review,Kuenzel1980} The detailed character of the defects leading to these lines is still unclear but they are frequently observed in MBE-grown GaAs layers of lesser quality.

Well separated from the exciton peaks, recombinations of free electrons with acceptor-bound holes appear. They are marked in this spectrum as (eC), because they are ascribed to carbon atoms on As-sites. Other impurities are not detected. They would have been observed at slightly different but specific wavelengths as summarized in Ref.\cite{PL_review}

PL spectra taken with heterostructures containing a 2DEG show a much larger intensity compared with those from bulk undoped GaAs layers indicating that the PL originates from the interface region. We see little influence of the 2DEG on the position of the exciton line which is in agreement with earlier observations\cite{Kundrotas2010} and the assumption that the PL is generated in the GaAs near the 2DEG.\cite{Shen1999} In a few samples we noticed that the (X) peak splits into two which could be a consequence of the 2DEG interacting with the exciton transition and might be connected with the H-band.\cite{Yuan1985} It seems that this feature is, however, not relevant for the analysis of the Ga quality.

PL from the AlGaAs would show up at much shorter wavelengths. Luminescence from the GaAs substrate or the smoothing superlattice was not seen, probably because the light is nearly completely absorbed in the 1 ${\mu}m$ thick GaAs buffer layer.

\section{Photo luminescence and 2DEG mobility}

MDSI structures are usually grown for low-temperature transport experiments, particularly in the regime of the integer and the fractional quantum Hall effect. Many MBE growers, including the authors,  produce MDSIs regularly for determining the mobility to control the status of the respective MBE-system. These samples are therefore very attractive for PL studies  because they are both readily available and comparable structures are grown repeatedly over long periods of time. 

 The significance of PL for the characterization of heterostructures is obvious from the data of  Fig.~\ref{Fig_4_1} which shows spectra of two different MDSI samples grown in the same MBE chamber during the same growth campaign. They only differ in the use of the Ga cells containing the lots Ga4 and Ga8, respectively. The structures of these two samples are identical, but they have very different transport properties. The mobility of D110131A is $9.96\cdot10^6 cm^2/Vs$ while, in contrast, the sample D110706E is nearly insulating. This discrepancy in the transport properties corresponds to an impressive qualitative difference between the PL spectra.
 
  Peaks due to the exciton recombination (X) are seen in both spectra at \mbox{1.5151 eV} but their intensities are very different. The broad peak (eC) at \mbox{1.4935 eV} in the D110706E spectrum signals the presence of carbon since it corresponds to free electron to neutral acceptor (carbon) transitions. It is obvious from Fig.~\ref{Fig_4_1} that the wafer with the high mobility has both a much larger intensity at the exciton peak and a much lower one related to the electron-carbon recombination compared to the low mobility one. This correlation between mobility and PL will be corroborated by many other spectra presented later in this publication.
  % The case of Fig. \ref{Fig_4_1} is, however particularly relevant because both samples are grown under exactly identical conditions, except for the gallium source, in the same MBE chamber.

  \begin{figure}
   \includegraphics[width=0.6\textwidth]{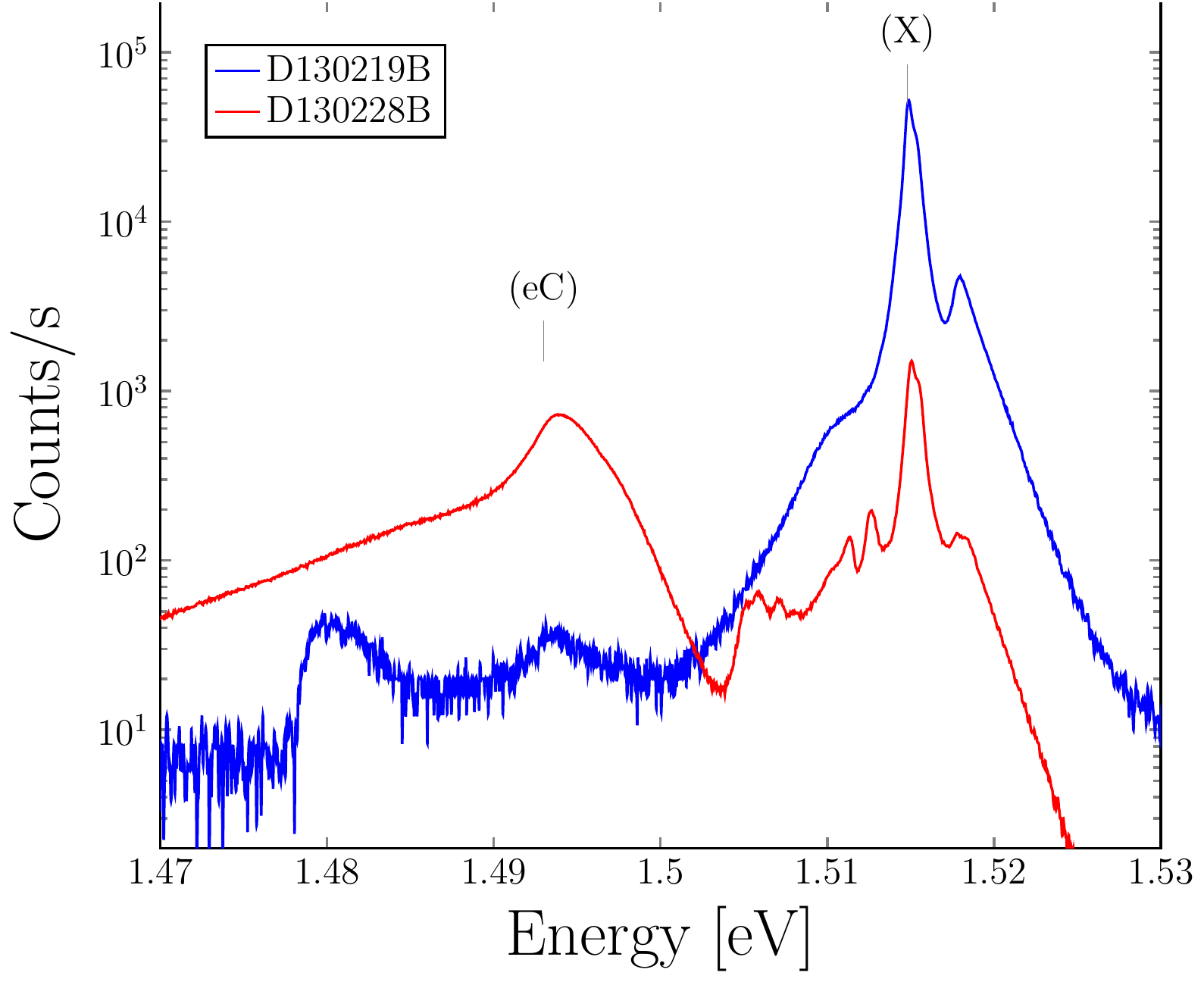}\\
   \textbf{\caption{\label{Fig_4_8}
    \textbf{PL spectra  of the samples D130219B (\textit{Ga{\slshape 4}}) and D130228B (\textit{Ga8}) which are produced in the same growth campaign as those of Fig. \ref{Fig_4_1} but about two years later. The large discrepancy between the two samples in the (X) vs. (eC) ratio is nearly unchanged. 
    }}}
  \end{figure}

Both gallium ingots labeled Ga4 and Ga8 came from the same supplier, Geo Gallium.\footnote{The gallium used in this work was provided by Geo Gallium, Salindres, France, by MCP, Wellingborough, UK, and by UMC, Lyndhurst, NJ, USA.} Other ingots from both lots showed similar chemical analysis (GDMS) results and had similar residual resistance ratios. The ingots were packaged in the standard PE foil and were mounted in the MBE without any further treatment. However, the Ga8 lot with the lot number 0036474 was of the standard MBE grade while the Ga4 one (0309683) underwent two more recrystallisation steps in the factory.

To clarify the role of the gallium source for mobility and PL, we now discuss the correlation between the PL spectra and the transport properties of a large number of  MDSI structures. At first, we restrict ourself to samples from one growth campaign lasting several years in one MBE system labeled "D-chamber". This chamber has also been used for the samples of Fig.~\ref{Fig_4_1}. The PL spectra of Fig.~\ref{Fig_4_8} are obtained from wafers grown in this chamber two years later, nine days apart. The PL intensities are now plotted on a log scale. The exciton (X) line of the wafer using the  Ga4-source is again much larger than the one of the wafer using Ga8. Likewise, the carbon-acceptor recombination (eC) remains to be  much larger in the Ga8 than in the Ga4 wafer. However, the mobility of the wafer with the Ga4 increased to $13.4\cdot10^{6}\, cm^{2}/Vs$ while the one using the Ga8 is only $0.27\cdot10^{6}\, cm^{2}/Vs$.

\begin{table}
\centering\caption{Comparison of mobility with PL properties of MDSI samples grown in the  D-chamber using  gallium Ga4.}
\ra{1.3}
\begin{tabular*}{0.6\textwidth}{@{\extracolsep{\fill}}|c|c|c|c|@{}}\toprule \hline
%\begin{tabular}{0.75\textwidth}{@{\extracolsep{\fill}}|l|c|c|c|@{}}   \hline
	\textbf{Wafer} & \textbf{ $\mu$ } & \textbf{   (X) } & \textbf{   (X) / (eC)}\\   
 	&[10$^6$ ${cm^2}/{Vs}$] & \textbf{[counts/s]} & \\\midrule\hline
   	D101117A & 4.07& 30100  & 78\\   \hline
   	D110131A & 9.96& 47700  & 234\\   \hline
   	D110504B & 14.31& 25500  & 503\\\hline
   	D120228A & 8.82& 37100  & 481\\\hline
   	D120912A & 1.27& 23400  & 425\\\hline
   	D130219B & 13.38& 51000  & 1275\\\hline
   	D130409A & 9.22& 64100  & 704\\\hline
   	D130423B & 11.99& 28000  & 838\\\hline
   	D131105A & 10.31& 23800  & 513\\\hline
\bottomrule
\end{tabular*}
\label{Sample_D_Ga4}
\end{table}

\begin{table}
\centering\caption{Mobilities and PL properties of samples grown with gallium Ga8.}
\ra{1.3}
\begin{tabular*}{0.6\textwidth}{@{\extracolsep{\fill}}|c|c|c|c|@{}}\toprule \hline
%\begin{tabular}{0.75\textwidth}{@{\extracolsep{\fill}}|l|c|c|c|@{}}   \hline
	\textbf{Wafer} & \textbf{ $\mu$ } & \textbf{   (X) } & \textbf{   (X) / (eC)}\\   
 	&[10$^6$ ${cm^2}/{Vs}$] & \textbf{[counts/s]} & \\\midrule\hline
%\begin{tabular*}{@{}lccc@{}}\toprule
%	\textbf{Wafer} & \textbf{ $\mu$ } & \textbf{   (X) intensity   } & \textbf{   (X)vs.(eC)}\\ 
 %	&[10$^6$ $\frac{cm^2}{Vs}$] & \textbf{[counts]} & \textbf{ratio}\\\midrule
   	D101215C & 0.283 & 1400 & 1.2\\   \hline
   	D110706E &  n/a & 1641 & 1.2\\   \hline
   	D130228B & 0.267 & 1477 & 2\\    \hline
\bottomrule
\end{tabular*}
\label{Sample_D_Ga8}
\end{table}

In order to quantify the results of the PL measurements we define two criteria. The first one is the intensity of the exciton peak (X) at  \mbox{1.5151 eV}. Its intensity depends on the existence of non-radiative transitions via deep electronic levels in the GaAs. \cite{PL_review} These states provide additional non-radiative electron-hole recombination channels. Therefore a relatively small (X) PL intensity signals the existence of impurities or other defects. The chemical and (or) the crystallographic origin of these imperfections is usually not known.

\begin{figure}
 \includegraphics[width=0.6\textwidth]{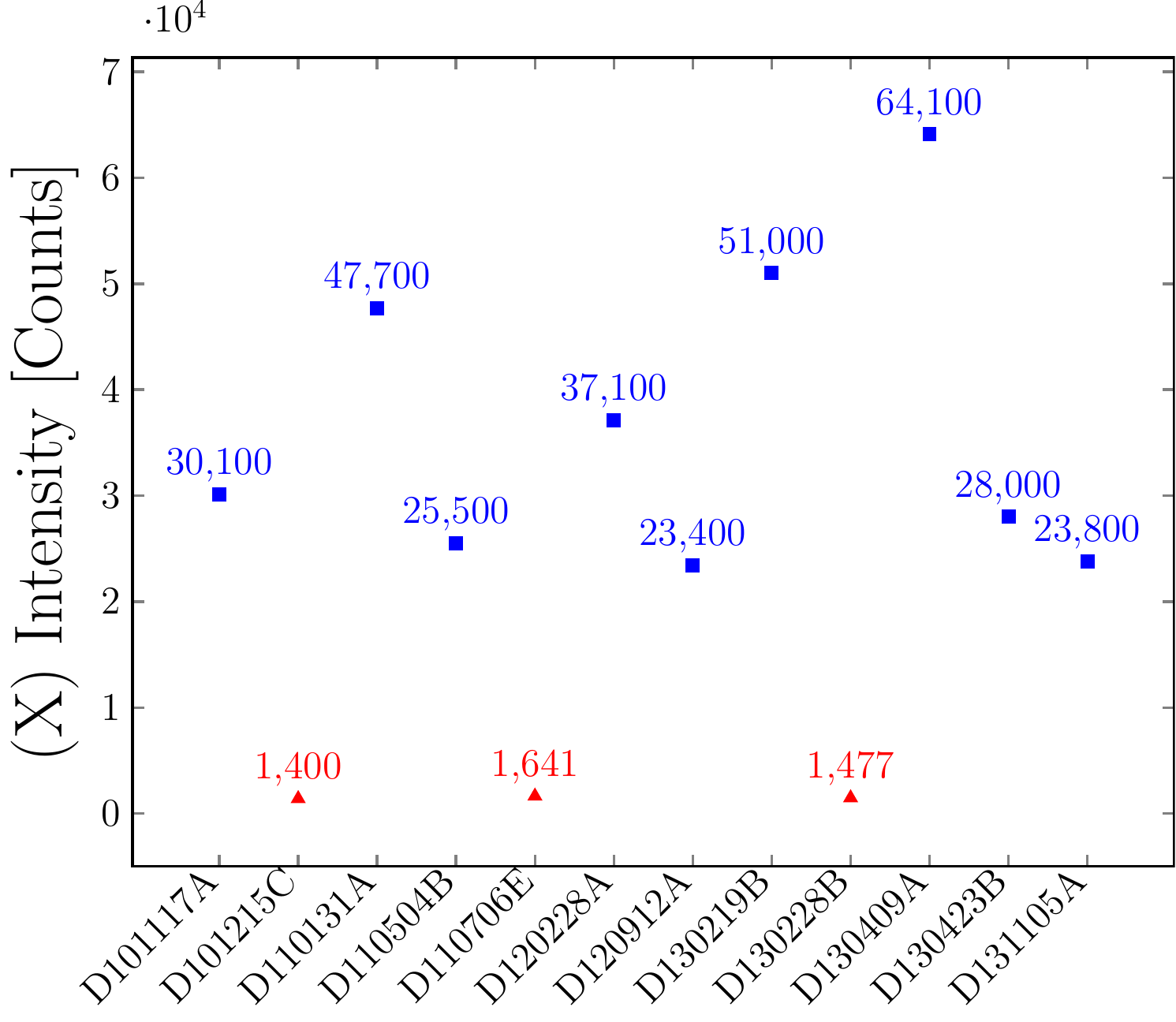}\\
 \textbf{\caption{\label{Fig_4_6}
  \textbf{Comparison of the exciton PL intensity for several wafers from one growth campaign. The samples marked with red triangles were grown with Ga8 , the other ones (blue squares) with Ga4.}}}
\end{figure}

\begin{figure}
 \includegraphics[width=0.6\textwidth]{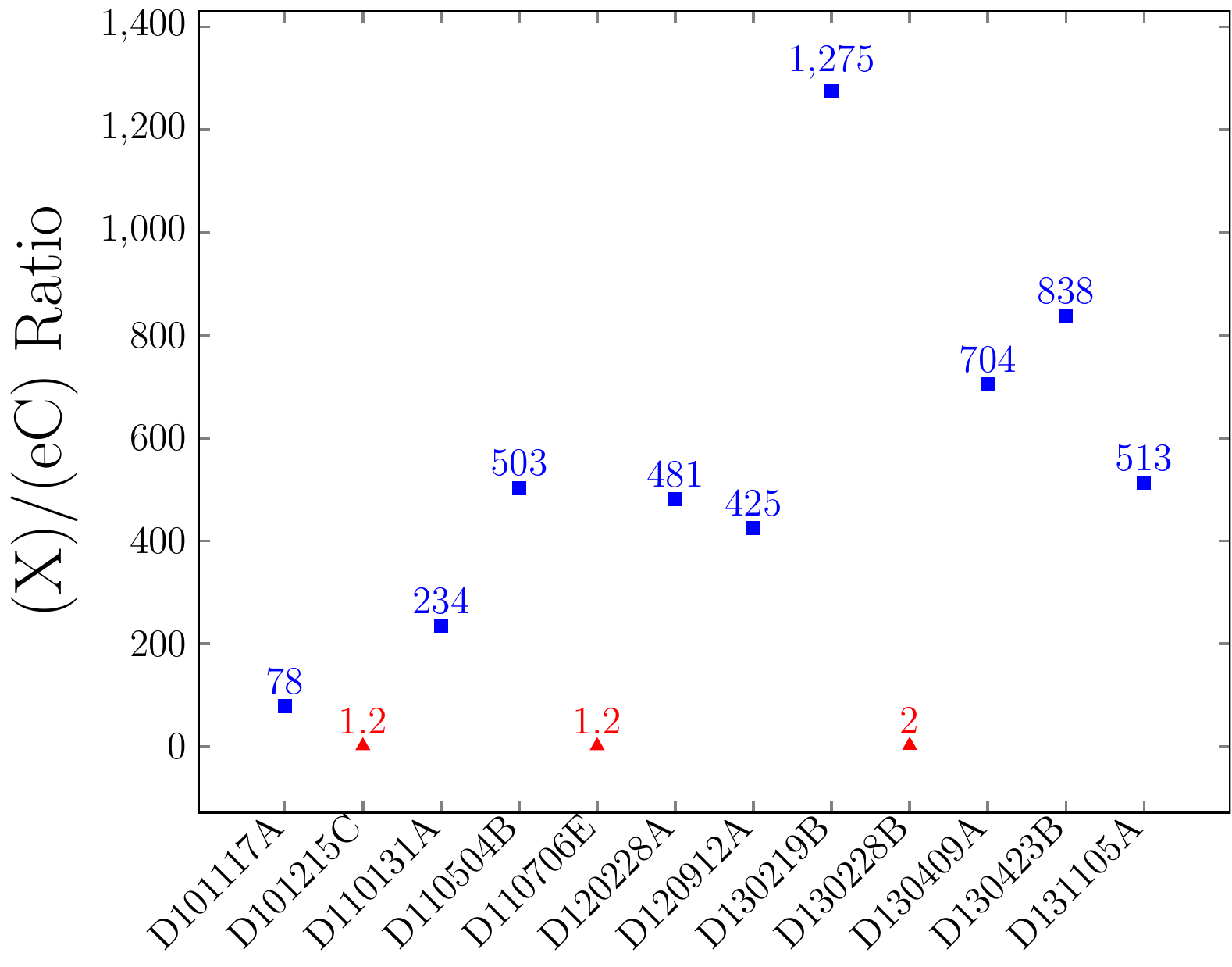}\\
 \textbf{\caption{\label{Fig_4_4}
  \textbf{(X) vs. (eC) peak intensity ratios of MDSI structures grown with Ga4 and Ga8, respectively. This ratio increases roughly with time if Ga4 (blue dots) is used, but remains very small with Ga8 (red triangles). }}}
\end{figure}

The second criterion is the intensity ratio between the exciton (X) and the acceptor peak (eC) at \mbox{1.493 eV}.
As Fig.s~\ref{Fig_4_1} and \ref{Fig_4_8} indicate, a large ratio is characteristic for high-mobility samples.
 The values of these criteria are compiled in tables \ref{Sample_D_Ga4} and \ref{Sample_D_Ga8} for our samples together with the respective mobilities and plotted in Fig.s \ref{Fig_4_6} and \ref{Fig_4_4}.

\begin{figure}
 \includegraphics[width=0.6\textwidth]{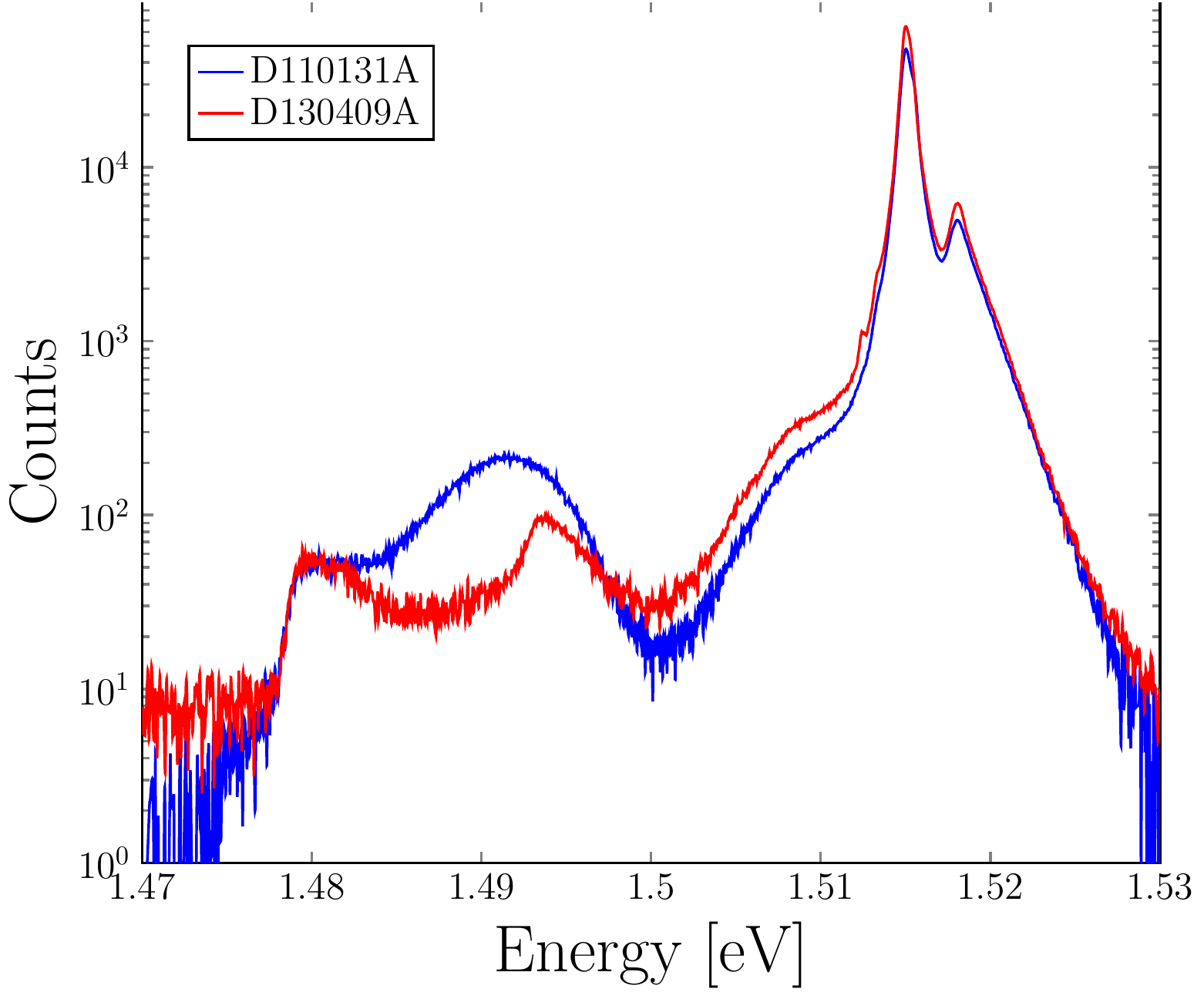}\\
 \textbf{\caption{\label{Fig_4_5}
  \textbf{PL spectra of two wafers grown about two years apart using the same "good" gallium lot Ga4. %with \textit{GEO MBE+} {\slshape 0309683} gallium. 
  The main change over this very long growth period is the reduction of the acceptor related luminescence.}}}
\end{figure}

The results for the (X)-peak intensities are shown in Fig.~\ref{Fig_4_6}. The blue dots refer to samples from wafers grown using Ga4, the red ones to those grown with Ga8. The difference is obvious. All samples grown with Ga4 have a much higher (X)-intensity than the ones grown with Ga8. This difference between the gallium sources is even more pronounced if one compares the (X)/(eC) ratio which is plotted in Fig. \ref{Fig_4_4}. This ratio is between about 100 and 1000 for Ga4-wafers while it is below 2 for the Ga8 ones. For the Ga4 samples one finds that the ratio tends to saturates about two years after commencement of growth. This could indicate a certain degree of "self-cleaning" of the Ga4 source. 

The effect of self cleaning is also visible in Fig. \ref{Fig_4_5} comparing PL spectra of wafers which are grown about two years apart using the same Ga source (Ga4). While the exciton intensity (X) is nearly unchanged within the 30 \% reproducibility margin, the PL due to the acceptor recombination (eC) is reduced in the wafer grown at the later time. The mobilities of the two wafers are similar, 9.96 and 9.22 [10$^6$ ${cm^2}/{Vs}$], respectively. For these specific wafers, the decrease of the (eC) PL does not corresponds directly with the mobility. This is an example that the correlation between the PL and the mobility is statistical.

\begin{figure}
 \includegraphics[width=0.6\textwidth]{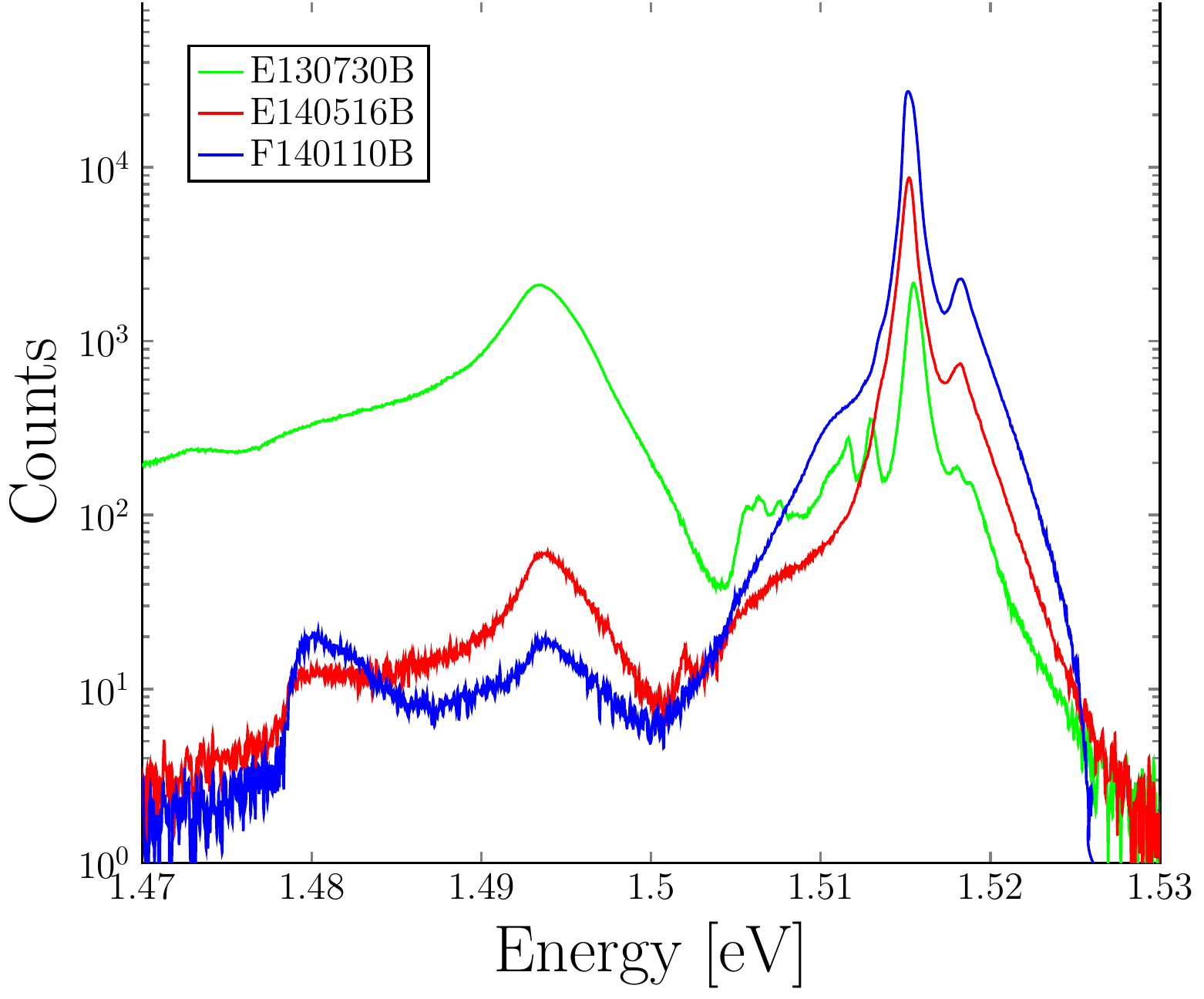}\\
 \textbf{\caption{\label{Fig_4_9}
  \textbf{PL spectra of wafers grown in two other MBE systems. The spectra shown in blue and red correspond to samples with reasonably high mobilities. The mobility of one in green is only $0.23\cdot10^6 {cm^2}/{Vs}$.}}}
\end{figure}

\begin{table}
\centering\caption{Comparison of MDSIs grown in different chambers using different Ga lots.}
\ra{1.3}
\begin{tabular*}{0.6\textwidth}{@{\extracolsep{\fill}}|c|c|c|c|@{}}\toprule \hline

%\begin{tabular}{0.75\textwidth}{@{\extracolsep{\fill}}|l|c|c|c|@{}}   \hline
	\textbf{Wafer} & \textbf{ Gallium} & \textbf{ $\mu$ [10$^6$ ${cm^2}/{Vs}$] } & \textbf{   (X) / (eC)}\\ \hline  
% 	&[10$^6$ ${cm^2}/{Vs}$] & \textbf{[counts/s]} & \\\midrule\hline
%\begin{tabular}{@{}lccc@{}}\toprule
%	\textbf{MDSI} & \textbf{Gallium} & \textbf{Mobility } & \textbf{(X)/(eC) ratio } %\\ 
% 	& & \textbf{[10$^6$ $\frac{cm^2}{Vs}$]} & \textbf{ [arb. units]}\\\midrule
%   	D130219B & GEO MBE+ 0309683 & 13.380 & 1300\\\hline
%   	D130228B & GEO 0036474 & 0.267 & 2\\   \hline
   	E130730B  & Ga4a & 0.23 &  1\\\hline
   	E140502A &  Ga4a & 0.70 & 4.7\\\hline
   	E140516B & Ga6  & 5.50 & 142\\\hline
   	F131121A & Ga4b & 4.41 & 242\\\hline
   	F140110B & Ga6 & 7.72 & 1405\\\hline
	F150122B & Ga6 & 13.75 & 1909\\\hline
   	F150130A & Ga6 & 1.14 & 1260\\\hline
   	C1070 & Ga2 & 6.96 & 228.49\\\hline
   	C1071 & Ga2 & 8.31 & 826.52\\ \hline
\bottomrule
\end{tabular*}
\label{MDSI_comparison}
\end{table}

\begin{table}
\centering\caption{List of gallium lots which is giving suppliers and lot number. The last column indicates the nominal purity, if extra crystallisation steps have been used and how the material was packaged.}
\ra{1.3}
\begin{tabular*}{1\textwidth}{@{\extracolsep{\fill}}|c|c|c|c|@{}}\toprule \hline
%\begin{tabular*}{1\textwidth}{@{}|l|c|c|c|@{}}   \hline
	\textbf{Ga label} & \textbf{Supplier} & \textbf{Lot Number } & \textbf{   Remarks)}\\ \hline  
	Ga2 & UMC & 811011 & 8N specification, foil \\\hline
	Ga4  & Geo & 0309683 & 7N, Extra purified, foil \\\hline
	Ga4a  & Geo & 0309683 & 7N, Extra purified,foil, unclear storage\\\hline
   	Ga4b  &Geo & 0309683 &  7N, Extra purified, Teflon bottle\\\hline
   	Ga6 & MCP  & GSC641A & 7N, Extra purified, Teflon bottle\\\hline
   	Ga8 & Geo & 0036474 & MBE grade, foil\\\hline

\bottomrule
\end{tabular*}
\label{Ga_list}
\end{table}

%\subsection{Relating electron mobility and PL in GaAs}

In table~\ref{MDSI_comparison} we compile mobility and PL data from wafers grown in three other MBE systems using Ga ingots originating from different lots and suppliers which are listed in table~\ref{Ga_list}. In Fig.~\ref{Fig_4_9} we show a selection of three spectra from these wafers corresponding to a wide range of mobilities. The gallium used for the E130730B, labeled Ga4a, is a different ingot from the same lot as the Ga4 used in the D-chamber. The one labeled Ga6 is stored in one Teflon bottle and needed to be liquefied, respectively.

\begin{figure}
 \includegraphics[width=0.6\textwidth]{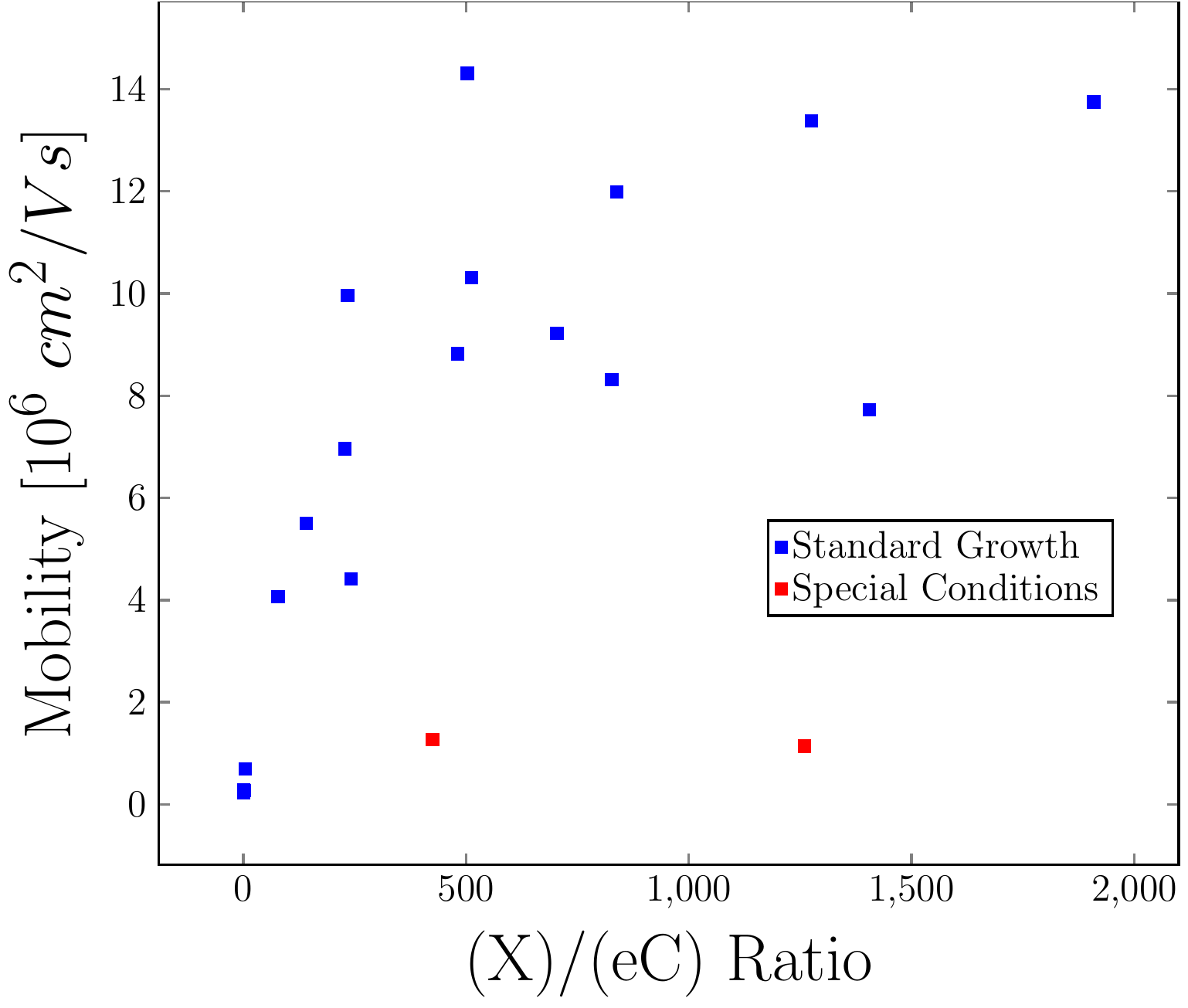}\\
 \textbf{\caption{\label{MobilityVsPL}
  \textbf{A plot of electron mobility vs. the (X)/(eC) ratio shows a correlation between the two quantities for many samples.}}}
\end{figure}

Generally, high mobility is related with good PL, i.e. a high exciton and a low acceptor-related emission. There are however, a few exceptions which will be discussed later. In Fig. \ref{MobilityVsPL} we plot the mobility values taken from the tables \ref{Sample_D_Ga4}, \ref{Sample_D_Ga8} and \ref{MDSI_comparison} vs the (X)/(eC) ratio. For most wafers (blue data ponts) there is a rough correlation between these two quantities. This correlation might be incidental considering that different impurities and processes influence the (X) and the (eC) intensities. The red points indicate samples which grossly deviate because different growth procedures or anomalous structures have been used.

One case is sample D120912A which has a good (X)/(eC) ratio of 433 but a low, 1.27 million $cm^2/Vs$ mobility. This wafer has been grown with an unusually high As cracker temperature. 
A high cracker temperature is not only a source of As$_2$ molecules but is also believed to be a source of impurities. It is unclear what the chemical nature of these impurities is, but apparently, they do not to enhance non-radiative recombination of excitons significantly. It had been suggested that  using a hot cracker leads to oxygen related impurities in the AlGaAs barrier layers\cite{Block1991} acting as deep acceptors. This appears to be less likely in our case because we do not observe any reduction in the electron concentration which would be a side effect of additional charge acceptors.

\begin{figure}
\begin {center}\includegraphics[width=0.6\textwidth]{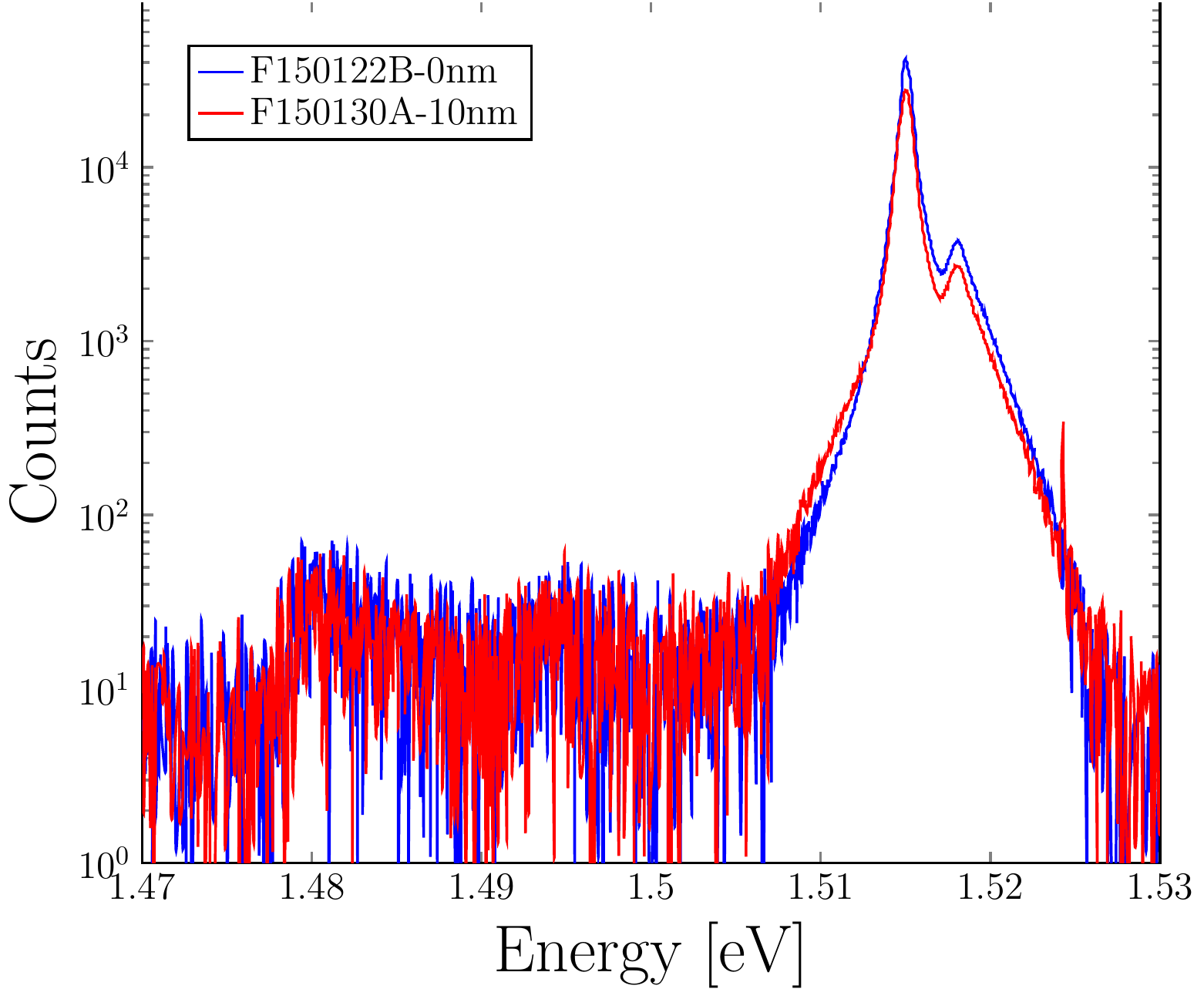}\end {center}
\protect\caption{\label{PLData}\bf{ PL data of a standard wafer and of one where Al impurities at 10~nm from the interface have been added to the GaAs.
Although the electron mobilities of the two samples differ by more than a factor of 10, the PL spectra are very similar.}}
\end{figure}

The origin of another deviation from the correlation between mobility and PL is much clearer than the one caused by the hot cracker. In that case, aluminum impurities have been added purposely to the GaAs by placing a thin layer of AlAs in the region where the 2DEG resides.\cite{Reichl2015} If located at the center of the wave function, i.e. 10 nm from the interface, the mobility degrades to 1.1  from the 13.7 million $cm^2/Vs$ of the reference sample without Al impurities. From the general behavior visible in Fig. \ref{MobilityVsPL} such a low mobility should also result in a large effect on the PL. This is clearly not the case as Fig. \ref{PLData} shows. There is very little difference in the PL between a reference spectrum without any Al in the GaAs and the one with an Al impurity layer.  Probably, the very short range of the local Al-impurity potential does not extend into the bulk region where the exciton recombination is expected to take place.\cite{Yuan1985}

It appears from Fig.~\ref{MobilityVsPL} as if the PL ratio saturates at high mobilities. This could mean that the mobility is no longer limited by the gallium. A final conclusion on this point is difficult, one reason is that the acceptor related luminescence (eC) becomes too small to be determined exactly, another one that the mobility of MDSI structures at 1 Kelvin does not seem to exceed 15$\cdot$10$^6$ ${cm^2}/{Vs}$ significantly. Thus, in the extremely high mobility regime one should better use double sided doped quantum wells to test extremely high mobilities and, at the same time, increase the precision of the PL in order to establish a better correlation in the extremely high mobility regime. In both MBE-systems studied here, mobilities between 25 and 30 million $cm^2/Vs$ were reached when those of the MDSIs are near 15 million  $cm^2/Vs$.

 \begin{figure}
 \begin {center}\includegraphics[width=0.6\textwidth]{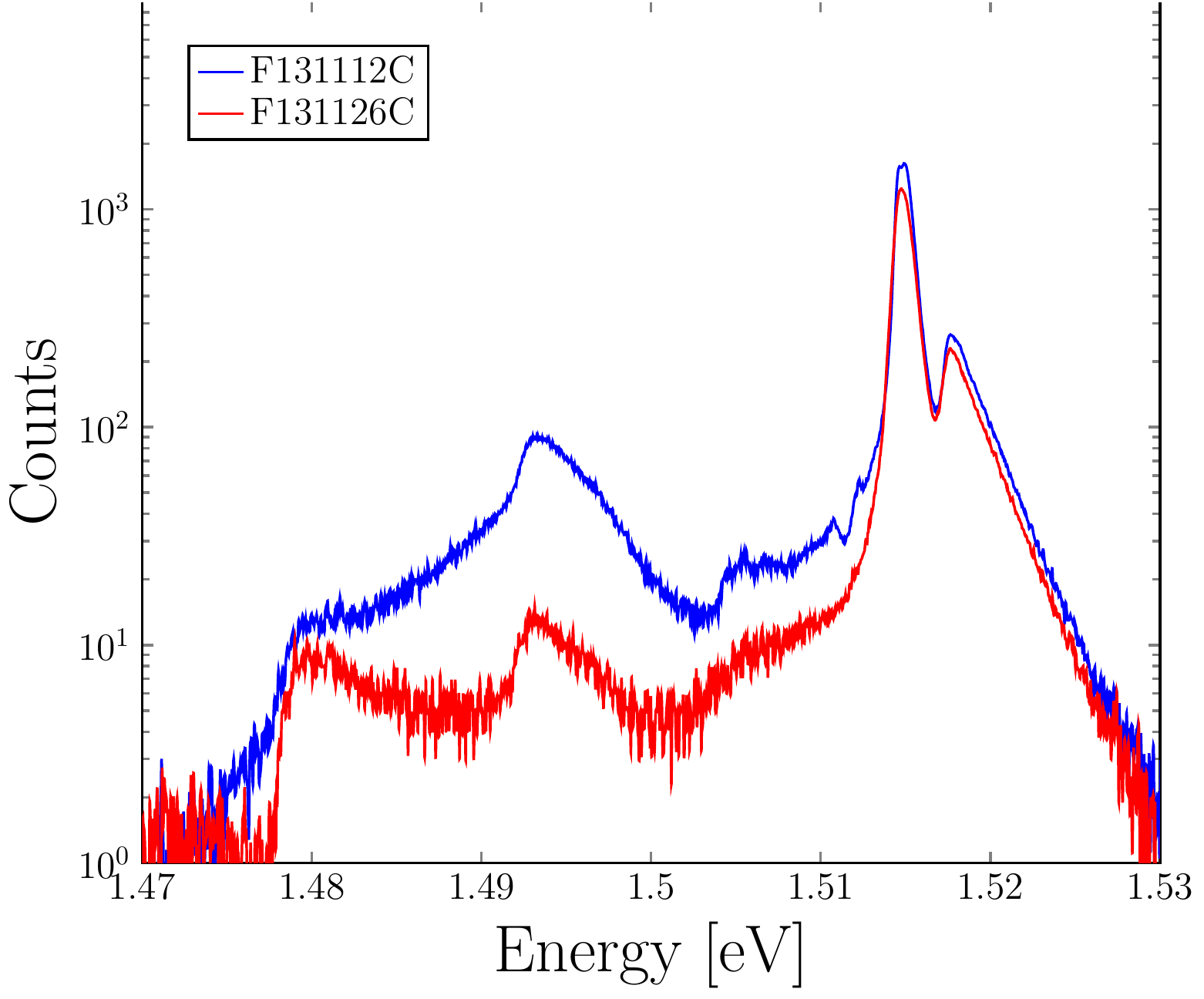}\end {center}
 \protect\caption{\label{PLBulk} \bf{PL spectra of "bulk" GaAs layers with 5 $\mu$m thickness showing that the PL signal is much weaker than in MDSI structures although the laser power density is increased to 2 $W/cm^2$. The red and the blue curves correspond to samples grown with Ga6 and Ga4b, respectively. The ratios between exciton and carbon luminescence are similar to those measured in MDSI structures grown in the same time period. The mobilities achieved with Ga6 are generally higher than those with Ga4b.}}
 \end{figure}

So far we restricted ourself to the discussion of the PL measured at MDSI structures because of their easy availability and because of the large PL signals. However, qualitatively similar results can also be obtained with relatively thick ($\approx$ 5 $\mu$m) undoped GaAs layers. In Fig. \ref{PLBulk} we show two spectra of bulk GaAs grown in the D-chamber with different Ga sources. The laser power density is increased to 2 $W/cm^2$. The (X)/(eC) ratios extracted from this measurement are similar to the ones obtained from MDSIs grown in the same time period. Thus, it is actually possible to assess the quality of a given Ga source soon after commencement of growth, even before the Al source is heated up to operation temperature.

\section{The role of aluminum}

The high reactivity of Al makes it likely that impurities are incorporated to a larger extend in the AlGaAs layer than in the GaAs. A possible candidate is oxygen which cannot be completely avoided in the residual gas of the MBE-vacuum chamber and exists in molecules like H$_2$O, CO$_x$, As$_2$O$_3$ etc.\cite{ACHTNICH1989} Oxygen impurities would probably act as deep acceptors and could be responsible for a weak electron scattering limiting the mobility at the presently observed maximal values.

It would be very attractive to study such impurities the in AlGaAs by the PL technique.\cite{Wicks81} Unfortunately, it is not straightforward to use MDSI structures for that purpose.  It turns out that the PL of the AlGaAs spacer layers is enhanced in the region of the Si doping layer and information about small impurity contents is difficult to obtain from the spectra. Thus undoped samples preferably with structures confining the excitons and enhancing the PL need to be grown for a study of the AlGaAs. In such samples, the desired direct comparison with mobility will be difficult, however. 

 On the other hand, it seems that in routine MBE operation there is no problem with the purity of the Al ingots, in contrast to the one of the Ga. As a test of the Al quality, we recently compared the effect of different Al lots on the mobility: One MBE system was loaded with two cells containing different Al ingots with residual resistance ratios of 8000 and 33000, respectively. No mobility differences have been found in the resulting MDSI structures grown on the same day. This indicates that there are no large  fluctuations in the electron mobility caused by a varying Al quality.

\section{Discussion}

We presented PL spectra of a large variety of samples which were all grown under nominally excellent MBE conditions. Nevertheless, the electron mobilities of these samples vary over nearly two orders of magnitude accompanied by significant variations of the respective PL spectra. Most impressive are the differences between wafers grown in the same MBE system using a "good" (Ga4) vs. a "bad" (Ga8) gallium source. With good gallium the intensity of the exciton PL (X) of the first wafers grown in a campaign is already high and does not increase with the number of samples produced. The electron mobility of these first wafers is already in the several million $cm^2/Vs$ regime, it increases during the campaign and reaching a maximum of about 15 million and more than 25 million $cm^2/Vs$ in double-sided doped quantum wells grown at about the same time. This increase is accompanied by a significant decrease of the PL from shallow impurities (eC). 

In contrast, the use of "bad" Gallium leads to wafers showing a small exciton (X) PL accompanied by a large one at (eC). Electron mobilities remain very small, even after an extended growth campaign during which PL does not change significantly. We did not, however, attempt to overheat this Ga by 100 or 200 K. With such a measure Gardner et al.\cite{Gardner2016} reported very recently that they succeeded to "turn" a poor starting material into a good one.

Compiling data from wafers grown in different MBE systems shows that most wafers show a rough correlation between the PL and the electron mobility. In the moderate mobility regime (< 10 million  $cm^{2}/Vs$) the ratio of the exciton emission (X) to the acceptor related luminescence (eC) increases with mobility. At higher values, the mobilities of the MDSI structures saturate at around 15 million $cm^{2}/Vs$ while the (X)/(eC) ratio continues to grow. Thus, it is possible that in this regime the mobilities are no longer limited by the purity of the Ga source.

Exceptions to the correlation between mobility and PL are observed in anomalous structures. For example a monolayer of AlAs atoms does not change the PL significantly although it is detrimental to the mobility. This is probably a consequence of the fact that the PL does not directly originate from the 2DEG region but from the neighboring bulk. Another case of an anomalous behavior, i.e. a low mobility but a good PL, is observed with a high cracker temperature of the As source. The chemical nature of the impurities is not known in this case but we  suspect that they are located mainly in the AlGaAs and therefore do not affect the GaAs PL.

Concerning the quality of the Ga, the main conclusion of this study is that if the gallium is "good" then deep impurity levels are more or less absent from the beginning. Shallow ones (carbon) are initially present but tend to reduce with the number of growth runs. The shallow ones may originate from the gallium, the arsenic, the residual gas of the MBE system, or from all of them. If the Ga is "bad", then non-radiative processes are very strong and do not reduce during a standard growth campaign. Mobilities remain at unsatisfactory low levels.

Greater care in the manufacturing process is a factor. In the case of the Ga4 and the Ga6 lots, the respective suppliers agreed to perform more recrystallization steps as usual. These Ga lots led indeed to both better electron mobilities and PL. 
%\footnote{One example is the Ga lot GEO 0309683 which was produced by the Geo company in cooperation with W.D. and W.W.} 

Unfortunately, it remains unclear what makes a gallium lot "bad". This cannot be only a manufacturing problem but must also be connected with storage and handling. We observe that ingots from the Ga4 lot can behave both either as good or as bad Ga. This is demonstrated by the data of the wafers E130730B and E140502A which are grown using a different ingot from the same lot as the Ga4. 

It has been noticed several years ago that wet-chemical etching of the ingots may improve the mobilities because it may remove contaminants at the surface of the ingots.\cite{CHAMBERS1986,SHAYEGAN1988} Possibly the plastic foil which is usually used for packaging either gives up material to the gallium, as suspected in these papers or the foil is not impenetrable to oxygen. Indeed, the gallium labeled Ga6 which led to the best PL is stored in a Teflon bottle which is kept in a separate air tight inert gas container.

In a previous experiment by Schmult et al.\cite{Schmult2009}, the oxygen contamination of the gallium was detected directly by analyzing the Ga vapor above an effusion cell with a mass spectrometer in a vacuum system which was equivalent to the one of an MBE. 
A significant amount of Ga$_2$O was detected in the molecular beam. This result is in agreement with an early conjecture that the gallium is a source of oxygen impurities in an MBE system.\cite{ACHTNICH1989} A direct correlation of this oxygen contamination with the mobilities or the PL is, however, still missing. One question is also why Ga$_2$O should be relevant at all, because oxides are known to desorb from the GaAs at the substrate temperatures at growth time. 

\section{Conclusions}

 It has been pointed out by L.N. Pfeiffer et al. that the continuous development of the vacuum technology has been crucial for the achievement of extremely high electron mobilities.\cite{2003Pfeiffer} Considering the scientific and financial efforts put into the technology it is frustrating that the choice of a bad Ga lot has such a catastrophic consequence. The comparison of the two lots, Ga4 and Ga8, in the same MBE growth campaign, is a particular dramatic example.

It is unsatisfactory that there is no easy way to judge the Ga quality before using it in the MBE. However, we showed that the PL technique is a tool which does not only provide more information about the impurities but can also be used as an early warning of Ga deficiencies. If the exciton intensity in early samples is small then it is very likely that the gallium is "bad" and that there is little probability of improvement with time. In such a case the gallium should be replaced if an alternative is available. If, on the other hand, the exciton intensity in early wafers is high, then the gallium quality should not be an obstacle to reach high mobilities. 

We could also show that certain Ga lots led to mobilities in the several million regime right with the first few wafer grown. This demonstrates that more effort by the suppliers in purification and packaging could be substantial for reaching mobilities close to the theoretical limit which is set to the order of 100 million $cm^2/Vs$ by Hwang and Das Sarma.\cite{Hwang2008} 

\section{Acknowledgement}

We gratefully acknowledge the technical assistance of Jessica Gm\"ur, Siegfried Heider and Marcel Sturzenegger. Some samples were grown by Thomas Tschirky, Matthias Berl and Maik Hauser (MPI Stuttgart), respectively. The PL apparatus was setup with the help of Thomas Feil. Preliminary PL measurements done in cooperation with Igor Kukushkin motivated this study. Eric Seguy from Geo Gallium and Malcolm Harrower from MCP both agreed to produce especially purified Ga lots which indeed led to the best results in this study. We gratefully acknowledge the financial support of the Swiss National Foundation (Schweizerischer Nationalfonds, NCCR "Quantum Science and Technology").

%\bibliography{Fabian}

\end{document}